\documentclass[12pt]{iopart}
\usepackage{epsfig}
\usepackage{setstack}

\def\al{{\it et al}}

\def\0{\varnothing}

\def\Tr{{\rm Tr}}

\begin{document}
\title{Phase transition in a non-conserving driven diffusive
system}
\author{M. R. Evans$\dag$, Y. Kafri$\ddag$, E. Levine$\ddag$, D. Mukamel$\ddag$}
\address{$\dag$Department of Physics and Astronomy, University
of Edinburgh, Mayfield Road, Edinburgh EH9 3JZ, United Kingdom.\\
$\ddag$ Department of Physics of Complex Systems, Weizmann
Institute of Science, Rehovot, Israel 76100.}
\date{\today}

\begin{abstract}
An asymmetric exclusion process comprising positive particles,
negative particles and vacancies is introduced. The model is
defined on a ring and the dynamics does not conserve the number of
particles. We solve the steady state exactly and show that it can
exhibit a continuous phase transition in which the density of
vacancies decreases to zero. The model has no absorbing state and
furnishes an example of a one-dimensional phase transition in a
homogeneous non-conserving system which does not belong to the
absorbing state universality classes.
\end{abstract}


\maketitle

One dimensional driven diffusive systems have been a subject of
extensive study in recent years~\cite{Zia,Mukamel00,Schutz01}. A
prototypical model for studying these systems is the asymmetric
exclusion process in which particles move stochastically with a
preferred direction and hard-core exclusion~\cite{EvansBlythe}. In
this model the local dynamics conserves the particles. Asymmetric
exclusion processes with a single species of particles do not
exhibit phase transitions on a ring geometry. On the other hand
open systems in which the particle number is not conserved at the
boundaries can exhibit a variety of phase
transitions~\cite{Mukamel00, Evans95}. Generalisations of these
models to more than one species of particles have shown that phase
transitions and long-range order do exist in these systems even on
a ring geometry~\cite{Mallick,DJLS,Evans98,LBR00,Rittenberg99}.

Generally speaking one-dimensional models that exhibit phase
transitions are either (a) of the asymmetric-exclusion-process
type with a drive and conserving bulk dynamics or (b) have
non-conserving bulk dynamics with one or more absorbing states.
The latter case is related  to the directed percolation or other
absorbing state universality classes~\cite{Haye}. It is therefore
of interest to find a one-dimensional model where the bulk
dynamics is not conserving that exhibits a phase transition
unrelated to the absorbing state universality classes.

In this letter we introduce such a model and solve for its steady
state exactly. The model is defined on a ring of $L$ sites where
each site can be occupied by either a positive ($+$) particle, a
negative ($-$) particle or a vacancy ($0$). The model evolves
through the following conserving rates:
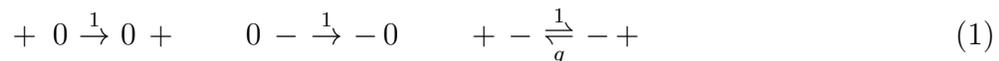
\begin{equation}
\label{eq:rates1} +\;0 \, \mathop{\rightarrow}\limits^{1} \, 0\,+
\qquad 0\,- \, \mathop{\rightarrow}\limits^{1} \, -\,0  \qquad
+\,- \, \mathop{\rightleftharpoons}\limits^{1}_{q} \, -\,+
\end{equation}
augmented by the following non-conserving rates
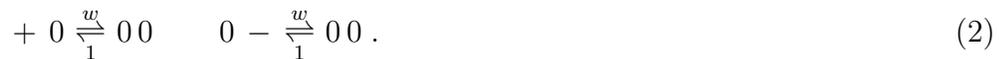
\begin{equation}
\label{eq:rate2} +\,0 \,
\mathop{\rightleftharpoons}\limits^{w}_{1}\, 0\,0 \qquad 0\,- \,
\mathop{\rightleftharpoons}\limits^{w}_{1} \, 0\,0 \;.
\end{equation}
Thus, the model generalizes the model of Derrida \al~\cite{DJLS}
and Arndt \al~\cite{Rittenberg99} through the introduction of the
process of creation and annihilation of particles. Note that any
configuration can be reached from any other except states with no
vacancies which are not dynamically accessible.

The rate $w$ controls the density of vacancies, denoted by
$\theta$, in the system. For large $w$ one expects the density of
vacancies to be high and for small $w$ it is expected to be low.
We will show that for $q<1$ the dependence of $\theta$ on $w$ is
not smooth and that for small enough $w$, $\theta$ is zero in the
thermodynamic limit. Thus the model exhibits a phase transition
from a ``fluid'' phase with a finite $\theta$ to a ``maximal
current'' phase where $\theta$ is zero (the nomenclature will be
explained below). The transition occurs at a critical value
$w_c>0$. The phase transition is found to be continuous with
\begin{equation}
\theta \sim |w-w_c|^\beta \;, \label{eq:th}
\end{equation}
where $\beta=1$. For $q>1$, on the other hand, the system is
always strongly phase separated~\cite{Evans98,LBR00,Rittenberg99}
with a single zero and two extensive pure domains of positive and
negative particles. Here there is no phase transition as $w$ is
varied.

A powerful technique for solving the steady-states of asymmetric
exclusion processes is the matrix product ansatz~\cite{DEHP}. This
involves representing the steady-state weights as the trace of a
product of matrices which depends on the microscopic
configuration. The matrices then obey certain algebraic rules
which are derived from the dynamics of the model. This technique
has yielded the exact behaviour at the various phase transitions
in many asymmetric exclusion models. The transitions solved within
the matrix product have been found to be robust for a large class
of systems~\cite{KrugSchutz}. However, in most models solved so
far using the matrix product, particle numbers are conserved in
the bulk (two exceptions are~\cite{Haye2} and~\cite{BEK}). In this
work we employ the matrix product technique to obtain an exact
solution for the steady state and phase transitions of the model
defined by (\ref{eq:rates1},\ref{eq:rate2}). It turns out that
despite the non-conserving dynamics of the present model we can
use the same matrices that have been previously used to solve
models with conserving dynamics.

We now proceed to outline the matrix product solution for the
steady state. The steady state weight of a configuration ${\cal
C}$ is represented by the trace of a product of matrices:
\begin{equation}
W_L({\cal C}) = \Tr \prod_{i=1}^{L} \left[ \delta_{\tau_i\,,\, +}\, D +
\delta_{\tau_i\,,\, -}\, E + \delta_{\tau_i\,,\, 0}\, A \right]\;,
\label{eq:matrixprod}
\end{equation}
where $L$ is the  system size and $\tau_j=+,-,0$ if site $j$ is
occupied by a $+,-$ or $0$ respectively. That is, a
matrix $D$ ($E$) represents a positive (negative) particle and $A$
represents a vacancy. It is easy to show using the technique
of~\cite{DEHP,DJLS} that $D,E$ and $A$ should satisfy
\begin{equation}
\label{eq:algebra} DE-qED = D+E \quad ; \quad   DA= AE = A \quad ;
\quad AA=wA
\end{equation}
to give the correct steady-state weights. These equations can be
satisfied by choosing
\begin{equation}
A=w\left|V\right>\left<V\right| \;, \label{eq:adef}
\end{equation}
where $\langle V| V \rangle=1$, $D |V \rangle =  |V \rangle$ and
$\langle V | E = \langle V |$. Then $D$, $E$ and $\langle V |$ are
identical to the matrices and vectors studied
in~\cite{Sasamoto99,Blythe00} where they are used to solve the
 single-species partially asymmetric exclusion
process with open boundaries.

We now wish to calculate the normalisation, i.e. the partition
function, which is given by the sum of the
weights~(\ref{eq:matrixprod}) of all accessible configurations:
\begin{equation}
{\cal Z}_L=\Tr\left[(A+D+E)^L-(D+E)^L\right]\; \label{eq:ZL}
\end{equation}
Using the form~(\ref{eq:adef}) of $A$ we first write the sum of
the weights of configurations with exactly $M$ vacancies on a
lattice of size $L$:
\begin{equation}
{\cal Z}_{L,M} = w^M \prod_{\mu=1}^{M} \sum_{n_\mu=0}^{\infty}
\langle V | C^{n_\mu} | V \rangle \; \delta_{ \sum_{\mu} n_\mu ,\;
L{-}M}
\end{equation}
where $C=D+E$. The sum over each $n_\mu$ corresponds to the
possible number of particles between two consecutive zeros. The
delta function enforces the constraint that the total number of
particles equals $L-M$ and the factor $w^M$ arises from the $M$
zeros. This form neglects the degeneracy in placing a given
configuration $\lbrace n_\mu \rbrace$ on a ring geometry, which is
bounded from above by $L$. It is straightforward to check using
bounds on the true partition function that this does not affect
any of the results present here.

In order to calculate the partition function ${\cal
Z}_L=\sum_{M=1}^{L} {\cal Z}_{L,M}$ it is convenient to replace
the delta function by a contour integral. Taking the limit of the
sum over $M$ to infinity yields
\begin{eqnarray}
{\cal Z}_L&=&\sum_{M=1}^{\infty} \oint\frac{dz}{2 \pi i} \frac{w^M
}{z^{L-M+1}} \prod_{\mu=1}^{M} \sum_{ n_\mu=0 }^{\infty} z^{n_\mu}
\langle V | C^{n_\mu} | V \rangle \label{eq:ZLint}\\
&=& \oint \frac{dz}{2 \pi i \; z^{L+1}}\left[
\frac{zwU(z)}{1-zwU(z)} \right] \;, \label{eq:int}
\end{eqnarray}
where $U(z)$ is defined as
\begin{equation}
U(z)=\sum_{n=0}^{\infty} z^n G_n \quad \mbox{with} \quad G_n
\equiv \langle V | C^n | V \rangle \;. \label{eq:U}
\end{equation}
The weight $G_n$ has been studied before. It is the
normalisation sum of the single species partially asymmetric
exclusion model on an open lattice of size $n$ and with particle
injection rate $1$ at the left boundary and removal rate $1$ at
the right boundary~\cite{Sasamoto99,Blythe00}.

Below, we shall consider
the density of vacancies
given by
\begin{equation}
\theta= \lim_{L \to \infty} \frac{\overline{M}}{L}= \lim_{L \to
\infty} \frac{w}{L}\frac{\partial \ln {\cal Z}_L}{\partial
w}\;, \label{eq:theta1}
\end{equation}
where $\overline{M}$ is the average number of vacancies in the
system. We shall also consider
the current
$J_L$ of positive particles (which is equal to that of the
negative particles). We define the current in the same way as for the
conserving system although here the density is not conserved. Thus
taking into account the inaccessibility of
configurations with no vacancies we find
\begin{eqnarray}
J_L&=&\frac{1}{{\cal Z}_L} \Tr \left[ \left( DA + DE-qED \right)
\left(C+A\right)^{L-2} -
\left(DE-qED \right) C^{L-2} \right]  \nonumber\\
&=& \frac{{\cal Z}_{L-1}}{{\cal Z}_{L}}\;, \label{eq:J}
\end{eqnarray}
where we have used the algebraic rules (\ref{eq:algebra}) and the
form (\ref{eq:ZL}) of ${\cal Z}_L$.

\noindent We now discuss the two distinct cases $q<1$ and $q>1$
separately:\\[1ex]

\noindent  {\it (i) The case of $q<1$}

Here the normalisation ${\cal Z}_L$
can be evaluated for large $L$ from the
integral (\ref{eq:int})   by
the saddle point method.  This method amounts to working in the grand
canonical ensemble. The term in the square brackets of (\ref{eq:int})
is then just the grand canonical partition function and ${\cal Z}_L
\sim (z^*)^{-L}$ where $z^*$ is the saddle point value of $z$,
i.e. the fugacity.  Thus, in the thermodynamic limit, we see from
(\ref{eq:J}) that $J_L \to z^* \equiv J$ which identifies the particle
current in the system as the fugacity. Also, we see from
(\ref{eq:theta1}) that
the density of vacancies is
given by
\begin{equation}
\theta =
 -w \frac{\partial \ln z^*}{\partial w} \;.\label{eq:theta}
\end{equation}
Thus one may identify $\theta$ as the order parameter and
$-\ln z^* = -\ln J$ as the analogue of the free energy density.

The saddle point equation for the integral
(\ref{eq:int}) may be written as
\begin{equation}
L=\frac{1}{1-wzU(z)}\,\left[\frac{zU'(z)}{U(z)}+wzU(z) \right] \;.
\label{eq:saddle}
\end{equation}
For $L \to \infty$ this equation is satisfied either by $1-wzU(z)
\sim {\cal O}(1/L)$ or by the term inside the brackets being of
order $L$. To analyse which of the two scenarios pertains one
needs to consider the properties of the increasing function
$U(z)$. Note from (\ref{eq:U}) that the convergence of $U(z)$ is
determined by the large $n$ form of $G_n$. For $q<1$
this quantity is known~\cite{Sasamoto99,Blythe00} to behave for
large $n$ as
\begin{equation}
G_n \simeq \; a  \;\frac{K^{-n}}{n^{3/2}} \label{eq:Casympt}
\end{equation}
where
\begin{equation}
K=\frac{1-q}{4} \quad \mbox{and} \quad a=\frac{4}{\sqrt{\pi}}
\left[ \prod_{i=1}^{\infty} \frac{(1-q^i)^3}{(1+q^i)^4} \right]
\;. \label{eq:Casas}
\end{equation}
From (\ref{eq:Casympt}) one deduces that $U(z)$ converges for $z
\leq K$ and $U'(z)$ diverges as $z \to K$. Thus, when $1-wzU(z) =
0$ for some $z \leq K$ we solve the saddle-point equation
(\ref{eq:saddle}) by choosing $z^*$ so that $1-wz^*U(z^*)\sim
{\cal O}(1/L)$. On the other hand when $1-wzU(z)>0$ for all $z
\leq K$, the divergence must come from the term in the square
brackets of equation (\ref{eq:saddle}) and we need $z^*=K(1-{\cal
O}(1/L^2))$. This can be deduced by noting that $U'(z) \sim
|K-z|^{-1/2}$ for $z \to K$. Thus, in the thermodynamic limit $L
\to \infty$, $z^*$ increases to $K$ as $w$ is decreased to $w_c$,
given by
\begin{equation}
w_c=\frac{1}{KU(K)} \;. \label{eq:critw}
\end{equation}
Any further decrease in $w$ leaves the value of $z^*$ unchanged.
Using equation (\ref{eq:theta}) one can see that for $w>w_c$,
$\theta$ decreases as $w$ decreases while for $w<w_c$, $\theta=0$.
Using $U(z)-U(K) \sim |K-z|^{1/2}$ it is easy to show by expanding
(\ref{eq:saddle}) that $\theta \sim |w-w_c|$ as $w \searrow w_c$,
 recovering (\ref{eq:th}).

Since $z^*$ is the current of particles and it saturates at
$K=(1-q)/4$ for $w<w_c$ we refer to this phase as the maximal
current phase. We refer to the phase $w>w_c$ as the fluid since
the typical configuration is a disordered arrangement of $+,-$ and $0$s.\\[1ex]

\noindent  {\it (ii) The case of $q >1$}

\noindent For $q>1$ it is known~\cite{Blythe00} that the $n$
dependence of $G_n$ is given by
\begin{equation}
G_n \sim \; \left( \frac{q}{q-1} \right)^n q^{n^2/4} \;.
\label{eq:PASEP}
\end{equation}
In this case $U(z)$ diverges for all $z>0$ and we have to impose
cutoffs on the sums in  equation (\ref{eq:ZLint}). That is, we
write
\begin{equation}
{\cal Z}_L=\sum_{M=1}^{L} \oint\frac{dz}{2 \pi i} \frac{w^M
}{z^{L-M+1}} \left( \sum_{ n=0 }^{L-M} z^{n} G_n \right)^M
\;.\label{eq:int2}
\end{equation}
From (\ref{eq:PASEP}) it is clear that the dominant contribution
to the sum is when $M=1$ and $n=L-1$. Then $J_L={\cal
Z}_{L-1}/{\cal Z}_{L} \simeq (q-1)\,q^{-L/2-1}$. Thus, the current
of particles is exponentially small in the system size and the
density of vacancies is zero. This corresponds to a strongly phase
separated state with a single vacancy followed by an extensive
block of positive particles (to the right of the vacancy) followed
by an extensive block of negative particles (to the left of the
vacancy).\\[1ex]

Finally, we discuss the transition between $q<1$ and $q>1$. Using
equations (\ref{eq:Casympt}) and (\ref{eq:Casas}) it is easy to
show that as $q \to 1$ from below, $KU(K)$ tends to zero. Thus
using equation (\ref{eq:critw}) we see that $w_c$ diverges as $q
\to 1$. Therefore at the $q=1$ transition the system changes from
a maximal current phase $J=(1-q)/4$ for $q<1$ to a strongly phase
separated state with $J={\cal O}(q^{-L/2})$ for $q>1$. In both of
these phases the density of vacancies is zero. Therefore the
transition has no relation to the non-conserving dynamics and is
instead related to the reversal of the bias~\cite{Blythe00}.

It is interesting to make a comparison between the transition from
the fluid to the maximal current phase and the denaturation
transition in DNA where the two strands of the molecule unbind at
a certain temperature. In models of this
transition~\cite{DNA1,DNA2} one assigns a Boltzmann weight $w$ to
bound base pairs and a weight $K^{-n}/n^c$ to  unbound
segments of length $n$. Here $K$ is a non-universal constant
whereas $c$ is a universal exponent depending on the dimension and
self-avoidance properties of the unbound DNA. For example, using a
random walk model yields $c=3/2$ in three dimensions~\cite{DNA1}.
As temperature is raised and $w$ decreases there is an unbinding
transition where the fraction of bound base pairs, $\theta$,
vanishes. The DNA models and the present model can be related by
identifying vacancies with bound pairs and blocks of particles
with unbound DNA loops. Then $G_n$ corresponds to the weight of
DNA loop of length $n$ and the grand canonical partition functions
of the two systems are the same~\cite{com2}.

The difference between the systems is that in the DNA the exponent
$c=3/2$ is a result of the fact that the one-dimensional molecule
is embedded in three dimensions. However, for the model considered
here the exponent $3/2$ arises from truly one-dimensional
phenomena related to the currents flowing in the system. This can
be seen from the relation between $G_n$ and the current of the
single-species partially asymmetric exclusion process. More
explicitly, for such a process the current in a system of size $n$
is given by $G_{n-1}/G_n$. Thus, the $n$ dependence of $G_n$ is
related to the $n$ dependence of the current in a single species
system of size $n$. In the context of the two species model we
have studied here such a system corresponds to an uninterrupted
block of particles, of length $n$, bounded between two vacancies.
This picture has recently been used to study the conditions for
phase separation in conserving models~\cite{conj}.

The model can easily be generalised to contain two additional
parameters $\alpha$ and $\beta$ by modifying the rates
(\ref{eq:rates1}) to read
\begin{equation}
+\;0 \, \mathop{\rightarrow}\limits^{\beta} \, 0\,+ \qquad 0\,- \,
\mathop{\rightarrow}\limits^{\alpha} \, -\,0 \qquad +\,- \,
\mathop{\rightleftharpoons}\limits^{1}_{q} \, -\,+
\end{equation}
and the rates (\ref{eq:rate2}) to
\begin{equation}
+\,0 \, \mathop{\rightleftharpoons}\limits^{\beta w}_{1} \, 0\,0
\qquad 0\,- \, \mathop{\rightleftharpoons}\limits^{\alpha w}_{1}
\, 0\,0 \;.
\end{equation}
Then one can take $A=w|V \rangle \langle W|$ where $\beta D |V
\rangle = |V \rangle$ and $\alpha \langle W|E=\langle W|$. This
generalisation should allow for a richer phase diagram than that
presented here.

Finally, we point out that the transition from the fluid to the
maximal current phase is lost if particles are not conserved
inside a block consisting only of particles. For example, a
process where $+ \to 0$ regardless of its neighbours destroys the
maximal current phase.

\ack The support of the Israeli Science Foundation is gratefully
acknowledged. MRE thanks the Einstein Center for support during a
visit to the Weizmann Institute.


\vspace{2cm}

\begin{thebibliography}{999}

\bibitem{Zia} B. Schmittmann and R. K. P. Zia in {\it Phase Transitions and
Critical Phenomena}, Vol. 17, Eds. C. Domb and J. L. Lebowitz,
(Academic Press, 1995).

\bibitem{Mukamel00} For a recent review see D. Mukamel in {\it Soft and Fragile Matter:
Nonequilibrium Dynamics, Metastability and Flow}, Eds. M.E. Cates
and M.R. Evans (Institute of Physics Publishing, Bristol, 2000);
{\tt cond-mat/0003424}.

\bibitem{Schutz01} G. M. Sch\"utz in {\it Phase Transitions and
Critical Phenomena}, Vol. 19, Eds. C. Domb and J. L. Lebowitz,
(Academic Press, 2001).

\bibitem{EvansBlythe} M. R. Evans and R. A. Blythe, Physica A in press, {\tt cond-mat/0110630}.

\bibitem{Evans95} M. R. Evans, D. P. Foster, C. Godr\`eche, and D.
Mukamel, Phys. Rev. Lett. {\bf 74} 208 (1995).

\bibitem{Mallick} K. Mallick, J. Phys. A: Math. Gen. {\bf 29},
5375 (1996).

\bibitem{DJLS} B. Derrida, S. A. Janowsky, J. L. Lebowitz and E. R.
Speer, J. Stat. Phys. {\bf 73}, 813 (1993).

\bibitem{Evans98} M. R. Evans, Y. Kafri, H. M. Koduvely, and D. Mukamel,
Phys. Rev. Lett. {\bf 80}, 425 (1998); Phys. Rev. E {\bf 58} 2764
(1998).

\bibitem{LBR00} R. Lahiri, S. Ramaswamy, Phys. Rev. Lett. {\bf 79}, 1150 (1997);
R. Lahiri, M. Barma, S. Ramaswamy, Phys. Rev. E {\bf 61}, 1648
(2000).


\bibitem{Rittenberg99} P. F. Arndt, T. Heinzel, V. Rittenberg, J. Phys. A
{\bf 31}, L45 (1998); J. Stat. Phys. {\bf 97}, 1 (1999).

\bibitem{Haye} H. Hinrichsen, Adv. Phys. {\bf 49}, 815 (2000).

\bibitem{DEHP} B. Derrida, M. R. Evans, V. Hakim, V. Pasquier, J.
Phys. A: Math. Gen. {\bf 26}, 1493 (1993).

\bibitem{KrugSchutz} J. S. Hager, J. Krug, V. Popkov and G. M. Sch\"utz,
Phys. Rev. E {\bf 63}, 056110 (2001) and references therein.

\bibitem{Haye2} H. Hinrichsen, S. Sandow and I. Peschel, J. Phys. A: Math. Gen. {\bf 29} 2643 (1996).

\bibitem{BEK} R. A. Blythe, M. R. Evans, Y. Kafri, Phys. Rev.
Lett. {\bf 85} 3750 (2000).

\bibitem{Sasamoto99} T. Sasamoto, J. Phys. A: Math. Gen. {\bf 32} 7109 (1999).

\bibitem{Blythe00} R. A. Blythe, M. R. Evans, F. Colaiori, and F. H. L. Essler, J.
Phys. A: Math. Gen. {\bf 33} 2313 (2000).

\bibitem{DNA1} D. Poland and H. A. Scheraga, J. Chem. Phys.
{\bf 45}, 1456 (1966); J. Chem. Phys. {\bf 45}, 1464 (1966).

\bibitem{DNA2} Y. Kafri, D. Mukamel and L. Peliti, Euro. Phys. J.
B. in press; {\tt cond-mat/0108323}.

\bibitem{com2} To make the correspondence more explicit one has to
identify $U$ and $V$ of the DNA literature with $U-1$ and
$wz/(1-wz)$ of the present paper.

\bibitem{conj} Y. Kafri, E. Levine, D. Mukamel, G. M. Sch\"utz, J.
T\"or\"ok, {\tt cond-mat/0204319}.

\end{thebibliography}
\end{document}